\documentclass[letterpaper, 10 pt, conference]{ieeeconf}  

\IEEEoverridecommandlockouts                              

\usepackage{mathtools}
\usepackage{amsmath, amssymb}
\usepackage{nicematrix} 
\usepackage{cases} 
\usepackage{changepage} 
\usepackage{standalone}
\usepackage{cite}
\usepackage{microtype} 
\usepackage{multirow}

\usepackage{amsthm}

\makeatletter
\newtheoremstyle{bfnote}%
    {\baselineskip} 
    {\baselineskip} 
    {\itshape} 
    {}
    {\bfseries}
    {.}
    {.5em}
    {\thmname{#1}\thmnumber{ #2}\thmnote{ (#3)}} 

\renewenvironment{proof}[1][\proofname]{\par
    \pushQED{\qed}%
    \normalfont \topsep6\p@\@plus6\p@\relax
    \trivlist
    \item\relax
    {\normalfont\bfseries 
    #1\@addpunct{.}}\hspace\labelsep\ignorespaces
    }{%
    \popQED\endtrivlist\@endpefalse
    }
\makeatother
\theoremstyle{bfnote}
\newtheorem{theorem}{Theorem}
\newtheorem{assumption}[theorem]{Assumption}
\newtheorem{remark}[theorem]{Remark}
\newtheorem{definition}[theorem]{Definition}
\newtheorem{lemma}[theorem]{Lemma}

\usepackage{algorithm}
\usepackage{algpseudocode}

\usepackage{tikz}
\usetikzlibrary{calc,matrix,shapes,arrows,patterns,intersections}
\usepackage{pgfplots}
\pgfplotsset{compat=newest}
\tikzset{
    axis break gap/.initial=1mm
}
\newlength\figH
\newlength\figW
\definecolor{matlab_blue}{rgb}{0 0.4470 0.7410}
\definecolor{matlab_yellow}{rgb}{0.9290 0.6940 0.1250}
\definecolor{matlab_purple}{rgb}{0.4940 0.1840 0.5560}
\definecolor{matlab_orange}{rgb}{0.8500 0.3250 0.0980}
\definecolor{matlab_red}{rgb}{1 0 0}
\definecolor{matlab_green}{rgb}{0 1 0}

\DeclarePairedDelimiter\norm{\lVert}{\rVert}%
\newcommand{\ie}{\textit{i}.\textit{e}.,}
\DeclareMathOperator{\nullspace}{null}
\DeclareMathOperator{\range}{range}
\DeclareMathOperator{\rank}{rank}

\DeclareMathAlphabet{\mymathbb}{U}{bbold}{m}{n}

\title{\LARGE \bf
NISE-PE Constraint: Data-Driven Predictive Control\\with Persistence of Excitation
}

\author{Lucca Heinze Faro, Yuanbo Nie, and Paul Trodden
\thanks{This work was supported by the MSc Conversion Scholarship of the (former) Department of Automatic Control \& Systems Engineering, University of Sheffield.}
\thanks{The authors are with the School of Electrical and Electronic Engineering, University of Sheffield, S1 3JD, United Kingdom. (Email: {\tt \{lheinzefaro1, y.nie, p.trodden\}@sheffield.ac.uk}.)}%
}

\begin{document}
\maketitle
\thispagestyle{empty}
\pagestyle{empty}

\begin{abstract}
Persistence of excitation (PE) is an important requirement for the successful operation of data-driven predictive control, as it ensures that the input--output data contains sufficient information about the underlying system dynamics. Nonetheless, this property is usually assumed rather than guaranteed. This paper introduces a novel data-driven predictive control formulation that maintains PE. The technical development that allows this is the characterisation of the \emph{nonexciting input set} (NIS), \ie~the set of inputs that lead to loss of PE, and the consequent derivation of a pair of disjoint, linear inequality constraints on the input, termed \emph{NIS exclusion PE}  (NISE--PE) constraint, that, if satisfied, maintain PE. When used in a predictive control formulation, these constraints lead to a mixed-integer optimal control problem with a single binary variable or, equivalently, a pair of disjoint quadratic programming problems that can be efficiently and reliably solved. Numerical examples show how these constraints are able to maintain PE during the controller's operation, resulting in improved performance over conventional approaches for both time-invariant and time-varying systems.
\end{abstract}

\section{Introduction}

Data-driven predictive control (DDPC) replaces dynamics models traditionally used in predictive control by a nonparametric representation using past input--output trajectories of the system~\cite{coulson_data-enabled_2019}. This formulation has seen a surge in interest in recent years as it allows the design of controllers directly from data, without an identification step~\cite{markovsky_behavioral_2021}, and notable progress has been made in stability, robustness and constraint satisfaction~\cite{berberich_overview_2025}. A core ingredient to the success of DDPC is the requirement that past input--output data can accurately capture the system dynamics. This can be ensured if the input satisfies a property known as \emph{persistence of excitation} \cite{willems_note_2005}. 

In traditional model-based predictive control, PE is enforced by embedding suitable constraints either in the main optimal control problem (OCP) or in an auxiliary OCP~\cite{genceli_new_1996, lu_robust_2023, verheijen_recursive_2022, aggelogiannaki_multiobjective_2006, marafioti_persistently_2014, zacekova_persistent_2013, hernandez_vicente_stabilizing_2019}. Many of these formulations could, in principle, be carried over to a DDPC setting. However, most rely on nonlinear matrix inequalities and thus produce nonlinear, nonconvex optimisation problems. To avoid nonconvexity, researchers have introduced approximations or alternative reformulations: linearised relaxations that are typically conservative~\cite{genceli_new_1996, lu_robust_2023}, and trace-based formulations~\cite{verheijen_recursive_2022} that recast the PE requirement and lead to a quadratically-constrained quadratic program (QCQP). While such QCQPs can often be solved quite efficiently in practice, they remain fundamentally nonconvex. Hence, although applicable to DDPC, these approaches either introduce nonconvexity or rely on approximations to avoid it.

For DDPC controllers, it is often assumed that a fixed input--output sequence, with a persistently exciting input, is obtained beforehand and sufficiently characterises the system dynamics for all time~\cite{coulson_data-enabled_2019,berberich_data-driven_2021-1, bongard_robust_2023}. In the practice, however, systems being controlled are typically nonlinear or time-varying, and thus the data may need to be updated online to capture the changing dynamics. In this scenario, PE is also typically assumed, rather than enforced~\cite{verhoek_data-driven_2021,berberich_linear_2022-1,baros_online_2022}. Hence, guaranteeing persistence of excitation online for DDPC is a relevant research direction~\cite{berberich_linear_2022-1,baros_online_2022}, as it can ensure that the data captures the local system dynamics at all times.

In the context of data-driven control, online input design methods that achieve a related property known as generalised PE \cite{markovsky_behavioral_2021} using exclusion constraints on the input are developed in \cite{van_waarde_beyond_2022,camlibel_shortest_2025}. Although these constraints can be used to construct input--output sequences able to parametrise all trajectories of a system, their direct inclusion in a DDPC formulation would result in a nonlinear and nonconvex optimisation problem even when, otherwise, the OCP would be a convex quadratic programming (QP) problem.

In this paper, a similar approach to \cite{camlibel_shortest_2025} is adopted, but focused on the problem of maintaining, rather than creating, persistence of excitation. Specifically, we consider a setting where a prior exciting input–output sequence is available, and we wish to maintain PE of the input–output sequence as new data are added. The first contribution is the characterisation of the \emph{nonexciting input set}, \emph{i.e.} the input selections that would lead to a loss of persistence of excitation. This concept is analogous to the affine set used to characterise the exclusion constraint in \cite[Lemma 5]{camlibel_shortest_2025}. The characterisation then facilitates the development of a novel pair of disjoint linear constraints, termed NISE-PE constraint, that, if satisfied, guarantees that the input lies outside the nonexciting input set and thus maintains PE. Subsequently, we formulate a persistently exciting DDPC controller that employs these constraints. Our approach involves solving, at each time step, \emph{up to} two quadratic programming problems that ensure the control input is persistently exciting. 

The rest of the paper is organised as follows. Section~\ref{sec:prob} defines the problem statement. Section~\ref{section: Nonexciting Input Set} develops the formulation of the novel PE constraints. In Section~\ref{sec:rank}, a persistently exciting data-driven predictive controller is formulated, employing these constraints. Section~\ref{sec:example} applies the PE-DDPC formulation to example systems, before conclusions are drawn in Section~\ref{sec:conc}.

\textbf{Notation:} For a vector $z_k^{} \in \mathbb{R}^q$, $k \in \mathbb{Z}$, we denote
\begin{equation*}
    z_{[i,j]} = 
    \begin{bNiceMatrix}
        z_{i} & \ldots & z_{j}
    \end{bNiceMatrix} \in \mathbb{R}^{(j-i+1)q}, \quad j \geq i, \quad i,j \in \mathbb{Z}.
\end{equation*} 
The $n$-length vector of ones is $\mymathbb{1}_n$, and the $n \times n$ identity matrix is $\mathbb{I}_n$. The Kronecker product is denoted by $\otimes$. The Hankel matrix of a signal ${u}_{\scriptstyle [k-T+1,k]}^{}$, $u_i \in \mathbb{R}^m$, is given by
\begin{equation}
    \label{eq:Hankel_matrix}
    \mathcal{H}_{\scriptscriptstyle L}^{} ({u}_{\scriptstyle [k-T+1,k]}^{})
    \coloneqq \begin{bNiceMatrix}
        u_{\scriptstyle k-T+1}^{}& \cdots & u_{\scriptstyle k-L+1}^{} \\
        \vdots &  & \vdots \\
        u_{\scriptstyle k-T+L}^{}& \cdots & u_{\scriptstyle k}^{}
    \end{bNiceMatrix}.
\end{equation}

\section{Problem Statement}
\label{sec:prob}


\begin{subequations} \label{eq:state_space_linear}
    Consider the linear, discrete-time system
    \begin{align}
        x_{k+1}^{} &= A x_k^{} + B u_k^{}, \\
        y_k^{} &= C x_k^{} + D u_k^{},
    \end{align}
    where $(x_k^{}, u_k^{}, y_k^{}) \in \mathbb{R}^n \times \mathbb{R}^m \times \mathbb{R}^p$ are the state, input and output at time $k$, $(A,C)$ is observable, and $(A,B)$ is controllable. This paper focuses on the DDPC problem of tracking a setpoint $({u}_{}^{S},{y}_{}^{S})$ while satisfying the constraints
\end{subequations}
\begin{equation}
    u_k^{} \in \mathbb{U}, \quad y_k \in \mathbb{Y} \quad \forall k.
\end{equation}

\begin{subequations} \label{eq:data_driven_MPC_entire_formulation}
    A standard DDPC problem, adapted from \cite{berberich_data-driven_2021-1}, is given by
    \begin{alignat}{3}
        & \min_{\substack{ \alpha, \, \sigma,  \\ \Bar{u}_{[-n,N-1]}^{}, \\ \Bar{y}_{[-n,N-1]}^{}}} & & & \begin{split} \sum_{i=0}^{N-1} \norm{\Bar{u}_{i}^{} - u_{}^{S}}_{R}^{2} + &\norm{\Bar{y}_{i}^{} - y_{}^{S}}_Q^{2}  \\ + \lambda_{\alpha} \norm{\alpha}_2^2 &+ \lambda_{\sigma} \norm{\sigma}_2^2 \end{split} \label{eq:data_driven_MPC_cost}\\
        & & \llap{\text{s.t.}} & &
        \begin{bNiceMatrix}
            \Bar{u}_{[-n,N-1]}^{} \\ \Bar{y}_{[-n,N-1]}^{} + \sigma
        \end{bNiceMatrix}
        &= 
        \begin{bNiceMatrix}
           \mathcal{H}_{\scriptscriptstyle N+n}^{} ({u}^{d})  \\ 
           \mathcal{H}_{\scriptscriptstyle N+n}^{} ({y}^{d})
        \end{bNiceMatrix} \alpha, \label{eq:data_driven_MPC_dynamics} \\
        & & & & \begin{bNiceMatrix}
            \Bar{u}_{[-n,-1]}^{} \\ \Bar{y}_{[-n,-1]}^{}
        \end{bNiceMatrix}
        & = 
        \begin{bNiceMatrix}
           {u}_{[k-n,k-1]}^{}  \\ {y}_{[k-n,k-1]}^{}
        \end{bNiceMatrix}, \label{eq:data_driven_MPC_initial}\\
        & & & & \begin{bNiceMatrix}
            \Bar{u}_{[N-n,N-1]}^{} \\ \Bar{y}_{[N-n,N-1]}^{}
        \end{bNiceMatrix}
        & = 
        \begin{bNiceMatrix}
           \mymathbb{1} \otimes {u}_{}^{S}  \\ \mymathbb{1} \otimes {y}_{}^{S}
        \end{bNiceMatrix}, \label{eq:data_driven_MPC_terminal} \\
        & & & & \Bar{u}_{i}^{},\Bar{y}_{i}^{} \in \mathbb{U} \times \mathbb{Y}, & \quad i = 0, \ldots,N-1,
        \label{eq:data_driven_MPC_inputoutput}
    \end{alignat}
    where $(\alpha, \sigma, \Bar{u}_{[-n,N-1]}^{}, \Bar{y}_{[-n,N-1]}^{}) \in \mathbb{R}^{T-(N+n)+1} \times \mathbb{R}^{(N+n)p} \times \mathbb{R}^{(N+n)m} \times \mathbb{R}^{(N+n)p} $ are decision variables and $(u_{}^{d},y_{}^{d}) \in \mathbb{R}^{mT} \times \mathbb{R}^{pT}$ are input--output data of length $T$ obtained in an offline experiment. The input applied to \eqref{eq:state_space_linear} at time $k$ is $u_k^{} = \Bar{u}_{0}^{*}$, where $\Bar{u}_{[-n,N-1]}^{*}$ is the solution to \eqref{eq:data_driven_MPC_entire_formulation}.
\end{subequations}

In this formulation, the system dynamics are represented in \eqref{eq:data_driven_MPC_dynamics} by the Hankel matrices of $u^d$ and $y^d$. This formulation is based on the Fundamental Lemma, originally proposed in \cite{willems_note_2005}, which allows the direct representation of linear systems using only past input--output data. Since the predicted trajectories in \eqref{eq:data_driven_MPC_dynamics} have a length of $N+n$, the Fundamental Lemma requires that $u_{}^{d}$ is persistently exciting of order $L=N+2n$ according to Definition \ref{def:PE_rank} \cite{berberich_data-driven_2021-1}. This condition ensures that $(u^d,y^d)$ accurately represents the system dynamics.

\begin{definition}[Persistence of excitation \cite{willems_note_2005, de_persis_formulas_2020}]
    \label{def:PE_rank}
    The signal ${u}_{\scriptstyle [k-T+1,k]}^{}$ is persistently exciting of order $L$ if its Hankel matrix $\mathcal{H}_{\scriptscriptstyle L}^{} ({u}_{\scriptstyle [k-T+1,k]}^{})$ has rank $m L$.
\end{definition}

\begin{remark}
    As $\mathcal{H}_{\scriptscriptstyle L}^{} ({u}_{\scriptstyle [k-T+1,k]}^{})$ has $mL$ rows, Definition \ref{def:PE_rank} requires that the Hankel matrix has full row rank. This implies it must have at least $mL$ columns and thus a minimum input--output data window of $T \geq (m+1)L - 1$ \cite{de_persis_formulas_2020, markovsky_behavioral_2021}.
\end{remark}

A relevant aspect of \eqref{eq:data_driven_MPC_entire_formulation} is that $(u_{}^{d}, y_{}^{d})$ are collected offline and fixed. Although, for simplicity, this paper considers the linear, time-invariant system \eqref{eq:state_space_linear}, the practical reality is that \eqref{eq:state_space_linear} merely models a nonlinear or time-varying system. In such cases, $(u_{}^{d},y_{}^{d})$ may need to be updated online to represent the changing dynamics. However, \eqref{eq:data_driven_MPC_entire_formulation} does not guarantee that the input signal is persistently exciting, making the enforcement of PE a necessary concern. The particular aims of this paper are thus to (i) propose novel linear PE constraints and, based on them, (ii) propose a persistently exciting DDPC algorithm that maintains PE given prior persistently exciting input data.


\section{NISE-PE Constraint Formulation}
\label{section: Nonexciting Input Set}

\subsection{Properties of the Input Hankel Matrix}

Before deriving the novel PE constraints, some properties of the Hankel matrix must be established, to the end of characterising  the \emph{nonexciting input set}. Lemma \ref{lemma:Hankel_matrix} provides necessary results for these derivations. 

For simplicity, the Hankel matrix \eqref{eq:Hankel_matrix} is recast as 
\begin{equation}
    \mathcal{H}_{\scriptscriptstyle L}^{} ({u}_{\scriptstyle [k-T+1,k]}^{})
    \coloneqq \begin{bNiceMatrix}
        H_{\scriptstyle 11, k}^{} & H_{\scriptstyle 12, k}^{} \\
        H_{\scriptstyle 21, k}^{} & u_{\scriptstyle k}^{}
    \end{bNiceMatrix},
    \label{eq:Hankel_matrix_simplified}
\end{equation}
where $H_{\scriptstyle 11, k}^{} \in \mathbb{R}^{m(L-1)\times (T-L)}, H_{\scriptstyle 12, k}^{} \in \mathbb{R}^{m(L-1)}$, and $H_{\scriptstyle 21, k}^{} \in \mathbb{R}^{m\times (T-L)}$ depend on the past signal ${u}_{\scriptstyle [k-T+1,k-1]}^{}$, but not on $u_k^{}$.

\begin{lemma}
    \label{lemma:Hankel_matrix}
    Suppose that, at time $k-1$, $u_{\scriptstyle [k-T,k-1]}$ is persistently exciting of order $L$. Then, at time $k$, the following properties hold for $u_{\scriptstyle [k-T+1,k]}$ and its Hankel matrix \eqref{eq:Hankel_matrix_simplified}:
    \begin{align}
        \rank \begin{bmatrix} H_{\scriptstyle 11, k}^{} & H_{\scriptstyle 12, k}^{} \end{bmatrix} &= mL-m, \label{lemma:Hankel_matrix_first}\\
        mL - 1 \leq \rank \left(\begin{bmatrix} H_{\scriptstyle 11, k}^{} \\ H_{\scriptstyle 21, k}^{} \end{bmatrix} \right) &\leq mL, \\
        mL - 1 \leq \rank \left(\mathcal{H}_{\scriptstyle L}^{}(u_{\scriptstyle [k-T+1,k]}^{}) \right) &\leq mL.
    \end{align} 
\end{lemma}

Condition \eqref{lemma:Hankel_matrix_first} implies that $\begin{bmatrix} H_{\scriptstyle 11, k}^{} & H_{\scriptstyle 12, k}^{} \end{bmatrix}$ has full row rank. Moreover, the hypothesis in Lemma \ref{lemma:Hankel_matrix} indicates that the aim of the proposed formulation is to \emph{maintain} PE rather than create it where it previously did not hold. 

A consequence of Lemma \ref{lemma:Hankel_matrix} is that choosing $T = (m+1)L - 1$, i.e., the smallest value such that the Hankel matrix can have linearly independent rows, represents the most demanding version of the problem of maintaining PE. In that case, $\mathcal{H}_{\scriptscriptstyle L}^{} ({u}_{\scriptstyle [k-T+1,k]}^{})$ is square, 
$\rank \left(\begin{bmatrix} H_{\scriptstyle k}^{\scriptstyle 11} \\ H_{\scriptstyle k}^{\scriptstyle 21} \end{bmatrix} \right) = mL-1$
and PE is maintained if and only if~$u_k^{}$ is chosen so that all rows of the Hankel matrix are linearly independent.

If $T > (m+1)L - 1$, $\mathcal{H}_{\scriptstyle L}^{}(u_{\scriptstyle [k-T+1,k]}^{})$ has fewer rows than columns and linear independence can possibly be maintained without careful selection of $u_{\scriptstyle k}^{}$, albeit at the expense of more variables and constraints in \eqref{eq:data_driven_MPC_entire_formulation}. Nevertheless, both scenarios are considered in the following developments.

\subsection{Nonexciting Input Set}

Based on the properties of the Hankel matrix, we define the \emph{nonexciting input set} at time $k$ as
\begin{equation}
    \label{eq: nonexciting input set 1}
    \mathbb{U}_k^{\boldsymbol{-}} \coloneq \{ u_k^{} \in \mathbb{R}^m \mid \rank \left( \mathcal{H}_{\scriptstyle L}^{}({u}_{\scriptstyle [k-T+1,k]}^{}) \right) < mL \}.
\end{equation}
The rank condition in \eqref{eq: nonexciting input set 1} means linear independence among the rows of $\mathcal{H}_{\scriptstyle L}^{}({u}_{\scriptstyle [k-T+1,k]}^{})$ is lost and thus $\mathbb{U}_k^{\boldsymbol{-}}$ is the set of inputs that cause loss of PE at time $k$. To \emph{maintain} PE, the input $u_{\scriptstyle k}^{}$ must then satisfy the exclusion constraint \[ u_{\scriptstyle k}^{} \notin \mathbb{U}_k^{\boldsymbol{-}}.\]

A more useful characterisation for $\mathbb{U}_k^{\boldsymbol{-}}$ can be obtained by considering the problem
\begin{equation}
    \begin{bmatrix}
        H_{\scriptstyle 11, k}^{\top} & H_{\scriptstyle 21, k}^{\top} \\
        H_{\scriptstyle 12, k}^{\top} & u_k^{\top}
    \end{bmatrix}
    z = 0, \quad z \in \mathbb{R}^{mL},
    \label{eq:left_null_space}
\end{equation}
which has $z=0$ as its only solution if and only if the rows of $\mathcal{H}_{\scriptstyle L}^{}({u}_{\scriptstyle [k-T+1,k]}^{})$ are linearly independent. The NIS can thus be equivalently defined as
\begin{equation*}
    \mathbb{U}_k^{\boldsymbol{-}} \coloneq \{ u_k^{} \in \mathbb{R}^m \mid \exists z \neq 0 \text{ such that \eqref{eq:left_null_space} holds} \}.
\end{equation*}

Given that, at time $k$, $u_{\scriptstyle k}^{} = \Bar{u}_0^*$ is yet to be determined, $\mathbb{U}_k^{\boldsymbol{-}}$ can be characterised by splitting \eqref{eq:left_null_space} into two equalities:  
\begin{align}
    \begin{bmatrix}
        H_{\scriptstyle 11, k}^{\top} & H_{\scriptstyle 21, k}^{\top}
    \end{bmatrix}
    z = 0, \label{eq:loss_PE_condition_1} \\
    \begin{bmatrix}
        H_{\scriptstyle 12, k}^{\top} & u_k^{\top}
    \end{bmatrix}
    z = 0 \label{eq:loss_PE_condition_2}.
\end{align}
If a solution $z_{\scriptstyle k}^{*} \neq 0$ to \eqref{eq:loss_PE_condition_1} exists, then it is a solution to \eqref{eq:left_null_space} if and only if it is also a solution to \eqref{eq:loss_PE_condition_2}. Therefore $\mathbb{U}_k^{\boldsymbol{-}}$ can be characterised by (i) first determining the conditions for which a solution $z_{\scriptstyle k}^{*} \neq 0$ to \eqref{eq:loss_PE_condition_1} exists and (ii) then determining the values of $u_k^{}$ for which \eqref{eq:loss_PE_condition_2} holds with $z = z_{\scriptstyle k}^{*}$. The following lemma provides the results for the first step.

\begin{lemma}
    \label{lemma:dimensions_of_U}
    For the equality system \eqref{eq:loss_PE_condition_1} at time $k$,
    \begin{enumerate}
        \item if $\rank \begin{bmatrix} H_{\scriptstyle 11, k}^{\top} & H_{\scriptstyle 21, k}^{\top} \end{bmatrix} = mL$, then $z_{\scriptstyle k}^{*} \equiv 0$.
        
        \item if $\rank \begin{bmatrix} H_{\scriptstyle 11, k}^{\top} & H_{\scriptstyle 21, k}^{\top} \end{bmatrix} = mL-1$, then $z_{\scriptstyle k}^{*} = \gamma z_{P,k}^{*}$, $\gamma \in \mathbb{R}$, where $z_{P,k}^{*} \neq 0$ is a particular solution to \eqref{eq:loss_PE_condition_1}. 
    \end{enumerate}
\end{lemma}

Applying the results of Lemma \ref{lemma:dimensions_of_U} to \eqref{eq:loss_PE_condition_2}, the main theoretical result of the paper can be derived, which is the characterisation of the nonexciting input set.
\begin{theorem}
    \label{theorem:loss_of_excitation_set}
    At time $k$, if the past signal $u_{[k-T+1,k-1]}$ is such that 
    \begin{enumerate}
        \item $\rank \begin{bmatrix} H_{\scriptstyle 11, k}^{\top} & H_{\scriptstyle 21, k}^{\top} \end{bmatrix} = mL$, then persistence of excitation is maintained for all $u_k^{} \in \mathbb{U}$. Therefore, $\mathbb{U}_k^{\boldsymbol{-}} = \emptyset$.

        \item $\rank \begin{bmatrix} H_{\scriptstyle 11, k}^{\top} & H_{\scriptstyle 21, k}^{\top} \end{bmatrix} = mL-1$, then $\mathbb{U}_k^{\boldsymbol{-}}$ is given by
        \begin{equation} \label{eq:second_definition_loss_excitation}
            \mathbb{U}_k^{\boldsymbol{-}} \coloneq \{ u_{\scriptstyle k}^{} \in \mathbb{R}^m \mid a_{\scriptstyle k}^\top u_{\scriptstyle k}^{} +c_{k}^{} = 0\}  \subset \mathbb{R}^{m},
        \end{equation}
        where $a_{\scriptstyle k}^{} \in \mathbb{R}^{m}$ and $c_{\scriptstyle k}^{} \in \mathbb{R}$ are given by
        \begin{equation}
            \label{eq:calculate_a_c}
            \begin{bNiceMatrix}
                b_{\scriptstyle k}^{} \\ a_{\scriptstyle k}^{}
            \end{bNiceMatrix} = z_{P,k}^{*}, \hspace{0.5em} c_{k}^{} = b_{\scriptstyle k}^\top H_{\scriptstyle k}^{\scriptstyle 12}, \hspace{0.5em} b_{\scriptstyle k}^{} \in \mathbb{R}^{mL-m}.
        \end{equation}
    \end{enumerate}
\end{theorem}

\begin{remark}
     Similar to the exclusion constraint in \cite{camlibel_shortest_2025}, \eqref{eq:second_definition_loss_excitation} is a hyperplane of dimension $m-1$.
\end{remark}

\subsection{NISE-PE constraint}

\begin{subequations} \label{eq:PE_constraint_1}
    As in Lemma \ref{lemma:Hankel_matrix}, the DDPC formulation requires no special treatment for the first scenario of Theorem \ref{theorem:loss_of_excitation_set}. Only the second scenario is relevant, since an inappropriate choice of $u_k^{}$ may lead to a loss of excitation. Considering the representation of $\mathbb{U}_k^{\boldsymbol{-}}$ given by \eqref{eq:second_definition_loss_excitation}, a pair of disjoint, linear persistence of excitation conditions can be devised as
    \begin{align} 
        a_{k}^\top \Bar{u}_0^{} + c_{k}^{}  &\geq +\|a_{k}^{}\| \varepsilon, \label{eq:PE_constraint_1_above}\\ 
        \text{or } a_{k}^\top \Bar{u}_0^{} + c_{k}^{}  &\leq -\|a_{k}^{}\| \varepsilon, \label{eq:PE_constraint_1_below}
    \end{align}
    where $\varepsilon \in \mathbb{R}$ is a tuning parameter. Constraints \eqref{eq:PE_constraint_1} ensure that $u_k^{}$ is at least $\varepsilon$ distant from $\mathbb{U}_k^{\boldsymbol{-}}$ and thus maintains PE.
\end{subequations}

Considering the disjoint nature of \eqref{eq:PE_constraint_1}, a mixed-integer PE constraint termed \emph{NIS Exclusion PE} constraint that enforces either \eqref{eq:PE_constraint_1_above} or \eqref{eq:PE_constraint_1_below} via a single binary variable can be formulated as
\begin{equation}
    \label{eq:PE_constraint_2}
    (2v-1) (a_{k}^\top \Bar{u}_0^{} + c_{k}^{})\geq \|a_{k}^{}\| \varepsilon, \quad v \in \{0,1\}.
\end{equation}
A visualisation of this constraint is shown in Fig.~\ref{fig:2d_input_final_plot}. As $\varepsilon \rightarrow 0$, $\|a_{k}^{}\| \varepsilon \rightarrow 0$ and \eqref{eq:PE_constraint_2} is trivially verified for $\varepsilon =0$.

\begin{figure}[ht]
    \centering
    \footnotesize
    \includegraphics[]{Figures/nonlinear_constraint_and_nonexciting_input_set.tex}
    \caption{\ref{fig:2d_input_nonexciting_input_set} $\mathbb{U}_{k}^{\boldsymbol{-}}$, feasible region for \eqref{eq:PE_constraint_2} with \ref{fig:2d_input_constraint_feasible_above} $v=1$ and \ref{fig:2d_input_constraint_feasible_below} $v=0$ for $m=2$, $\varepsilon = 0.2$, $L=5$, $T=14$, some random initial input sequence and $\mathbb{U} = \{u: \|x\|_\infty \leq 1\}$. The distance of the disjoint, linear constraints from $\mathbb{U}_{k}^{\boldsymbol{-}}$ can be adjusted by tuning the parameter $\varepsilon$.} 
    \label{fig:2d_input_final_plot}
\end{figure}

\begin{remark} 
    \label{remark:intersection of sets}
    If, at time $k$, $\mathbb{U}_k^{\boldsymbol{-}}$ does not intersect $\mathbb{U}$, then there exists a $\varepsilon > 0$ such that \eqref{eq:PE_constraint_2} is verified $\forall u_k^{} \in \mathbb{U}$, and thus any $u_k^{} \in \mathbb{U}$ maintains excitation. 
\end{remark}


\section{Persistently Exciting DDPC}
\label{sec:rank}


Constraint \eqref{eq:PE_constraint_2} ensures \emph{maintenance} of PE, and thus its application to a DDPC framework requires an initial persistently exciting input sequence, as  described in Assumption \ref{assump: full_rank_previus_step}. The same assumption is present in other DDPC implementations \cite{coulson_data-enabled_2019,berberich_data-driven_2021-1, bongard_robust_2023}.

\begin{assumption}
    \label{assump: full_rank_previus_step}
    At $k=0$, $u_{[-T+1,-1]}^{}$ is PE of order $N+2n$.
\end{assumption}

Based on the DDPC formulation \eqref{eq:data_driven_MPC_entire_formulation}, on the NISE--PE constraint \eqref{eq:PE_constraint_2}, and on Assumption \ref{assump: full_rank_previus_step}, the following persistently exciting DDPC (PE--DDPC) is proposed

\begin{subequations} 
\label{eq:PE_data_driven_MPC_entire_formulation_1}
    \begin{alignat}{3}
        & \min_{\substack{ \alpha, \, \sigma, \, v,  \\ \Bar{u}_{\scriptscriptstyle[-n,N-1]}^{}, \\ \Bar{y}_{\scriptscriptstyle[-n,N-1]}^{}}} &&
        \begin{alignedat}{2}
            \sum_{i=0}^{N-1} \norm{\Bar{u}_{i}^{} - u_{}^{S}}_{R}^{2} &+ \norm{\Bar{y}_{i}^{} - y_{}^{S}}_Q^{2}  \\ 
            + \lambda_{\alpha} \norm{\alpha}_2^2 &+ \lambda_{\sigma} \norm{\sigma}_2^2 
        \end{alignedat} \label{eq:PE_data_driven_MPC_objective} \\
        & \;\; \quad \text{s.t.} &&
        \begin{bNiceMatrix}
            \Bar{u}_{\scriptscriptstyle [-n,N-1]}^{} \\ \Bar{y}_{\scriptscriptstyle [-n,N-1]}^{} + \sigma
        \end{bNiceMatrix} =
        \begin{bNiceMatrix}
           \mathcal{H}_{\scriptscriptstyle N+n}^{} ({u}_{\scriptscriptstyle[k-T,k-1]}^{})  \\ 
           \mathcal{H}_{\scriptscriptstyle N+n}^{} ({y}_{\scriptscriptstyle[k-T,k-1]}^{})
        \end{bNiceMatrix} \alpha, \label{eq:PE_data_driven_dynamics1} \\
        & && \begin{bNiceMatrix}
            \Bar{u}_{\scriptscriptstyle [-n,-1]}^{} \\ \Bar{y}_{\scriptscriptstyle [-n,-1]}^{}
        \end{bNiceMatrix} = 
        \begin{bNiceMatrix}
           {u}_{\scriptscriptstyle [k-n,k-1]}^{}  \\ {y}_{\scriptscriptstyle [k-n,k-1]}^{}
        \end{bNiceMatrix},  \\
        & &&  \begin{bNiceMatrix}
            \Bar{u}_{\scriptscriptstyle [N-n,N-1]}^{} \\ \Bar{y}_{\scriptscriptstyle [N-n,N-1]}^{}
        \end{bNiceMatrix} = 
        \begin{bNiceMatrix}
           \mymathbb{1} \otimes {u}_{}^{S}  \\ \mymathbb{1} \otimes {y}_{}^{S}
        \end{bNiceMatrix}, \\
        & && (2v-1) (a_{k}^\top \Bar{u}_0^{} + c_{k}^{})\geq \|a_{k}^{}\| \varepsilon, v \in \{0,1\},  \label{eq:PE_constraint_3_1} \\
        & && \Bar{u}_{i}^{},\Bar{y}_{i}^{} \in \mathbb{U} \times \mathbb{Y},  \quad i = 0, \ldots,N-1
    \end{alignat}
    where $a_k$ and $c_k$ come from \eqref{eq:calculate_a_c}. In addition to the NISE--PE constraint  \eqref{eq:PE_constraint_3_1}, the proposed PE-DDPC differs from \eqref{eq:data_driven_MPC_entire_formulation} in that the dynamic constraint \eqref{eq:PE_data_driven_dynamics1} uses the most recent measurements, rather than a fixed set of data $(u_{}^{d},y_{}^{d})$ collected offline.
\end{subequations}

\begin{subequations} \label{eq:constraint_crossing_check}
    As \eqref{eq:PE_data_driven_MPC_entire_formulation_1} has a single binary variable $v$, it can alternatively be solved by considering two disjoint DDPC problems, one for each value of $v$, and implementing the solution that minimises the cost over the two problems. These remain QP problems similar in size to \eqref{eq:data_driven_MPC_entire_formulation}, as \eqref{eq:PE_constraint_1_above} and \eqref{eq:PE_constraint_1_below} are linear constraints. Furthermore, as pointed out in Remark \ref{remark:intersection of sets}, constraint \eqref{eq:PE_constraint_3_1} is required only if $\mathbb{U}_{k}^{\boldsymbol{-}}$ and $\mathbb{U}$ intersect. This can be verified, prior to solving \eqref{eq:PE_data_driven_MPC_entire_formulation_1}, by solving the linear program
    \begin{align}
        J(u) = \min_{u \in \mathbb{U}} \,& | a_k^\top u + c_k |,
    \end{align}
    for which the solution is $u^* = 0$ if, and only if, $\mathbb{U}_k^{\boldsymbol{-}}$ and $\mathbb{U}$ intersect. If $\mathbb{U}_k^{\boldsymbol{-}}$ and $\mathbb{U}$ do not intersect, \eqref{eq:PE_data_driven_MPC_entire_formulation_1} can be solved at time $k$ omitting \eqref{eq:PE_constraint_3_1}, reducing the complexity of the OCP.
\end{subequations}

The structured application of all of the previous elements is presented in Algorithm \ref{alg:cap}.

\begin{algorithm}
    \caption{Persistently Exciting Data-Driven MPC}\label{alg:cap}
    \renewcommand{\algorithmicrequire}{\textbf{Offline:}}
	\renewcommand{\algorithmicensure}{\textbf{Online:}}
    \begin{algorithmic}[1]
        \Require
        \State Define $\varepsilon$
        \State Apply an initial control sequence $u_{[-T,-1]}$ that is exciting according to Assumption \ref{assump: full_rank_previus_step}.
        
        \Ensure $k=0$
        \State Measure $u_k^{}$ and $y_k^{}$.
        \State Calculate $a_k^{}$ and $c_k^{}$ according to \eqref{eq:calculate_a_c}.
        \State Verify, using \eqref{eq:constraint_crossing_check}, if $\mathbb{U}_k^{\boldsymbol{-}}$ intersects $\mathbb{U}$.
        \If{$\mathbb{U}_k^{\boldsymbol{-}}$ and $\mathbb{U}$ intersect}
            \State Compute the solution $\Bar{u}_{[-n,N-1]}^{*}$ to \eqref{eq:PE_data_driven_MPC_entire_formulation_1}.
        \Else
            \State Compute the solution $\Bar{u}_{[-n,N-1]}^{*}$  to \eqref{eq:PE_data_driven_MPC_entire_formulation_1} without \eqref{eq:PE_constraint_3_1}.
        \EndIf
        \State Apply $u_k^{} = \Bar{u}_{0}^{*}$.
        \State Set $k = k+1$ and go back to step 3.
    \end{algorithmic}
\end{algorithm}

\section{Illustrative example}
\label{sec:example}
In this section, Algorithm \ref{alg:cap} is applied to a modified version of the linearised four-tank system \cite{berberich_data-driven_2021-1} given by
\setlength{\arraycolsep}{1.5pt}
\begin{align*}
    x_{k+1}^{} &= 
    \begin{bNiceMatrix}
            0.921 & 0 & 0.041 & 0 \\
            0 & 0.918 & 0 & 0.033 \\
            0 & 0 & 0.924 & 0 \\
            0 & 0 & 0 & 0.937
        \end{bNiceMatrix} x_{k}^{}
        + \begin{bNiceMatrix}
            \delta(k) & 0.001 \\
            0.001 & 0.023 \\
            0 & 0.061 \\
            0.072 & 0
        \end{bNiceMatrix} u_{k}^{}, \\
    y_{k}^{} &= \begin{bNiceMatrix}
        1 & 0 & 0 & 0 \\
        0 & 1 & 0 & 0
    \end{bNiceMatrix} x_k,
\end{align*}
with initial state $x_0^\top = \begin{bNiceMatrix} 0.4 & 0.4 & 0 & 0 \end{bNiceMatrix}$. The objective of the controller is to track the setpoint
\begin{equation*}
    (u^S,y^S) = \left( \begin{bNiceMatrix}
        1.04 \\ 0.99 \
    \end{bNiceMatrix},
    \begin{bNiceMatrix}
        0.65 \\ 0.77
    \end{bNiceMatrix} \right).
\end{equation*}
The prediction and data horizons are $N = 75$ and $T = 300$. The weights in \eqref{eq:PE_data_driven_MPC_objective} are $ Q = 50 \cdot \mathbb{I}_p$, $R = 10^{-5} \cdot \mathbb{I}_m$, $\lambda_{\alpha} = 0.1$, and $\lambda_{\sigma} = 1000$. The input and output constraint sets are $\mathbb{U} = [-1, 1.5]^2$ and $\mathbb{Y} = \mathbb{R}^2$.

The proposed controller is evaluated for two scenarios involving the parameter $\delta(k)$ in the $B$ matrix: $\delta(k) = 0.017$ and $\delta(k) = 0.017 + k \cdot 10^{-6}$, modelling a parameter drift.

\subsection{Case 1: $\delta(k) = 0.017$}

The system was simulated for 50 initial input sequences of length $T$, 100 values of $\varepsilon$ and a simulation length of $N_s = 1000$. The initial input sequence was uniformly sampled  from the interval $[0,1]^2$ and used to provide an initial excitation to the system for $k \in [1,T]$; Algorithm 1 was then applied from the instant $T+1$. The values of $\varepsilon$ were linearly sampled from $[10^{-4}, 3 \cdot10^{-1}]$. For this system, values of $\varepsilon$ smaller than $10^{-4}$ resulted in numerical errors when solving \eqref{eq:loss_PE_condition_1}. 

Fig.~\ref{fig:example1_input_output} compares input--output trajectories obtained using \eqref{eq:data_driven_MPC_entire_formulation} to trajectories obtained by Algorithm \ref{alg:cap} for two values of $\varepsilon$. As expected for a time-invariant system, on average all controllers were able to track the references, though differences in behaviour can be observed. The controller based on \eqref{eq:data_driven_MPC_entire_formulation} presented varying degrees of steady-state offset, resulting in a wider spread of the outputs in Fig.~\ref{fig:example1_input_output}. Algorithm~\ref{alg:cap} eliminates this offset and reduced the shaded regions for the outputs at the expense of introducing oscillations around the setpoints and increasing variance for the input. 

This trade-off in performance can be verified in Table~\ref{tab:squared_tracking}, which presents the mean squared tracking errors for for the scenarios compared in Fig.~\ref{fig:example1_input_output}. Smaller values of $\varepsilon$ eliminate the offset with smaller increases in input oscillation, resulting in better input and output tracking. Higher values of $\varepsilon$ can further improve output tracking at the expense of a higher input tracking error.
\begin{figure}[ht]
    \centering\footnotesize 
    \begin{tabular}{cc}
        \input{Figures/LTI_case1_vs_case2_y1} & \input{Figures/LTI_case1_vs_case2_u1} \\
        \input{Figures/LTI_case1_vs_case2_y2} & \input{Figures/LTI_case1_vs_case2_u2} \\
        \input{Figures/LTI_case1_vs_case3_y1} & \input{Figures/LTI_case1_vs_case3_u1} \\
        \input{Figures/LTI_case1_vs_case3_y2} & \input{Figures/LTI_case1_vs_case3_u2}
    \end{tabular}
    \caption{Inputs and outputs for $\delta(k) = 0.017$ and \ref{example1_epsilon_0} \eqref{eq:data_driven_MPC_entire_formulation}, \ref{example1_epsilon_00698} \eqref{eq:PE_data_driven_MPC_entire_formulation_1} with $\varepsilon = 0.0698$ and \ref{example1_epsilon_03} \eqref{eq:PE_data_driven_MPC_entire_formulation_1} with  $\varepsilon = 0.3$. The dashed lines represent the average values at each time step, while the shaded regions represent the minimum and maximum values at each time step over all initial sequences.} 
    \label{fig:example1_input_output}
\end{figure}

\begin{table}[ht]
    \centering
    \caption{Mean squared tracking error over all initial input sequences over the interval $[N_s-T+1,N_s]$ for Case 1} 
    \begin{tabular}{c|ccc}
               & $\varepsilon=0$        & $\varepsilon=0.0698$   & $\varepsilon=0.3$ \\
        \hline
         $y_1$ & $1.3437 \cdot 10^{-6}$ & $3.6999 \cdot 10^{-7}$ & $2.8518 \cdot 10^{-7}$\\
         $y_2$ & $1.5160 \cdot 10^{-6}$ & $3.4010 \cdot 10^{-7}$ & $2.5051 \cdot 10^{-7}$ \\
         $u_1$ & $3.4253 \cdot 10^{-5}$ & $1.2517 \cdot 10^{-5}$ & $1.3959 \cdot 10^{-4}$\\
         $u_2$ & $3.7662 \cdot 10^{-5}$ & $1.5602 \cdot 10^{-5}$ & $1.4229 \cdot 10^{-4}$
    \end{tabular}
    \label{tab:squared_tracking}
\end{table}

Fig.~\ref{fig:example1_cond_vs_epsilon} evaluates the relation between $\varepsilon$ and the condition number of $\mathcal{H}_{\scriptstyle L}^{}({u}_{\scriptstyle [k-T+1,k]}^{})$. These plots consider the average, minimum and maximum condition numbers over all initial sequences for the final interval $[N_s-T+1,N_s]$, which reduces influences from the initial transient. It can be observed that, as the value of $\varepsilon$ increases, the condition number decreases, indicating a better-conditioned problem. 
\begin{figure}[ht]
    \centering
    \footnotesize
    \input{Figures/cond_vs_epsilon_LTI.tex}
    \caption{Steady-state condition number of $\mathcal{H}_{\scriptstyle L}^{}({u}_{\scriptstyle [k-T+1,k]}^{})$ for all initial input sequences over the interval $[N_s-T+1,N_s]$ as a function of $\varepsilon$.} 
    \label{fig:example1_cond_vs_epsilon}
\end{figure}

\subsection{Case 2: $\delta(k) = 0.017 + k \cdot 10^{-6}$}

In this scenario, the controlled dynamics correspond to a time-varying system. Although this contradicts the assumption of a time-invariant system, this example is used to show that the data update with a persistently exciting input can approximately capture the dynamics of the system and result in a good tracking performance. The parameters were kept as in Case 1, except for the simulation length, which was increased to $N_s = 6000$ to better illustrate the results.

Fig.~\ref{fig:example2_input_output} compares the system response when controlled using \eqref{eq:data_driven_MPC_entire_formulation} to its response to Algorithm~\ref{alg:cap}. As expected, the controller based on \eqref{eq:data_driven_MPC_entire_formulation} is not able to track any of the setpoints, as the initial data does not accurately capture the changing dynamics. The controllers based on Algorithm \ref{alg:cap} can track the outputs while the inputs drift away from the setpoint. This is expected as the specified setpoints no longer correspond to an equilibrium point for the system due to the parameter drift and thus the higher weight assigned to the outputs in the problem formulation promotes a better output tracking. 

Furthermore, as the data used to represent the system is collected while the dynamics are changing, the resulting nonparametric model can be seen as an interpolation over the collection period and thus is different from the dynamics of the controlled system. This results in the offset in output tracking when using Algorithm~\ref{alg:cap}.
\begin{figure}[ht]
    \centering\footnotesize
    \begin{tabular}{cc}
        \includegraphics[]{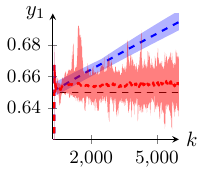} & \includegraphics[]{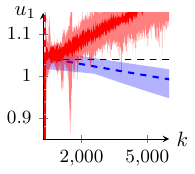} \\
        \includegraphics[]{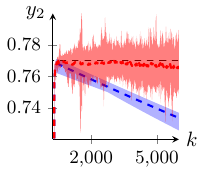} & \includegraphics[]{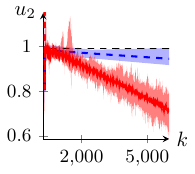} \\
        \includegraphics[]{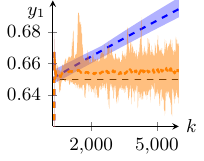} & \includegraphics[]{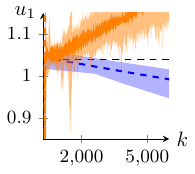}\\
        \includegraphics[]{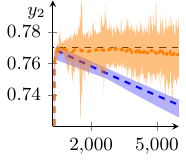} & \includegraphics[]{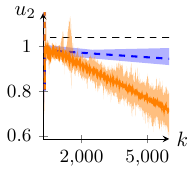} \\
    \end{tabular}
    \caption{Inputs and outputs for $\delta(k) = 0.017 + k\cdot 10^{-6}$ and and \ref{example1_epsilon_0} \eqref{eq:data_driven_MPC_entire_formulation}, \ref{example1_epsilon_00698} \eqref{eq:PE_data_driven_MPC_entire_formulation_1} with $\varepsilon = 0.0698$ and \ref{example1_epsilon_03} \eqref{eq:PE_data_driven_MPC_entire_formulation_1} with  $\varepsilon = 0.3$. . The dashed lines represent the average values at each time step, while the shaded regions represent the minimum and maximum values at each time step over all initial sequences.} 
    \label{fig:example2_input_output}
\end{figure}

\vspace{1\baselineskip}
\section{Conclusions and future work}
\label{sec:conc}
In this paper, the set of nonexciting inputs, which cause PE to be lost in a data-driven predictive control setup, was characterised. This nonexciting input set was then used to develop a novel set of disjoint, linear inequality constraints, termed NISE-PE constraint, that, at the current time, maintain PE of a previously exciting input sequence. These constraints were employed in a DDPC problem that ensures maintain persistence of excitation; it was shown that, despite its mixed-integer nature, the problem can be solved efficiently as two alternative quadratic programming problems. Future work will focus on comparing the proposed linear PE constraints to other nonlinear and linearised constraints used in the literature.

\section*{ACKNOWLEDGMENT}
We thank Bernardo A. Hernández Vicente for the helpful discussions and comments.



\bibliographystyle{IEEEtran}
\bibliography{root}

\appendix

\subsection{Proof of Lemma \ref{lemma:Hankel_matrix}}
    At time $k$, the Hankel matrix of inputs is
    \begin{equation*}
        \begin{aligned}
            \mathcal{H}_{\scriptscriptstyle L}^{} &({u}_{\scriptstyle [k-T+1,k]}^{}) \\
            &= \begin{bmatrix}
                u_{k-T+1}^{} & u_{k-T+2}^{} & \cdots & u_{k-L}^{} & u_{k-L+1}^{} \\
                u_{k-T+2}^{} & u_{k-T+3}^{} & \cdots & u_{k-L+1}^{} & u_{k-L+2}^{} \\
                \vdots & \vdots &  & \vdots \\
                u_{k-T+L-1}^{} & u_{k-T+L}^{} & \cdots & u_{k-2}^{} & u_{k-1}^{} \\
                u_{k-T+L}^{} & u_{k-T+L+1}^{} & \cdots & u_{k-1}^{} & u_{k}^{}
            \end{bmatrix} \\
            &= \begin{bNiceMatrix}
                H_{\scriptstyle 11, k}^{} & H_{\scriptstyle 12, k}^{} \\
                H_{\scriptstyle 21, k}^{} & u_{\scriptstyle k}^{}
            \end{bNiceMatrix}.
        \end{aligned}
    \end{equation*}
    
    Fig. \ref{fig:comparison_hankel_matrices} highlights the correspondences between the Hankel matrix at time $k$, $\mathcal{H}_{\scriptscriptstyle L}^{} ({u}_{\scriptstyle [k-T+1,k]}^{})$ and the Hankel matrix at time $k-1$, $\mathcal{H}_{\scriptscriptstyle L}^{} ({u}_{\scriptstyle [k-T,k-1]}^{})$. While the last $T-L$ columns of $\mathcal{H}_{\scriptscriptstyle L}^{} ({u}_{\scriptstyle [k-T,k-1]}^{})$, i.e. all of its columns except the first one, correspond to 
    \[\begin{bNiceMatrix}
        H_{\scriptstyle 11, k}^{} \\
        H_{\scriptstyle 21, k}^{}
    \end{bNiceMatrix},\]
    the last $mL-m$ rows of $\mathcal{H}_{\scriptscriptstyle L}^{} ({u}_{\scriptstyle [k-T,k-1]}^{})$ correspond to 
    \[\begin{bNiceMatrix}
        H_{\scriptstyle 11, k}^{} & H_{\scriptstyle 12, k}^{}
    \end{bNiceMatrix}.\]
    
    \begin{figure}[ht]
        \centering
        \stepcounter{figure}
        \begin{gather*}
            \begin{bNiceMatrix}[last-row,last-col=6]
                \centering
                u_{k-T}^{}     & u_{k-T+1}^{}   & \cdots & u_{k-L-1}^{} & u_{k-L}^{} & \\
                \vdots         & \vdots         &        & \vdots       & \vdots     & \color{matlab_red}H_{11,k}^{} \\
                u_{k-T+L-2}^{} & u_{k-T+L-1}^{} & \cdots & u_{k-3}^{}   & u_{k-2}^{} & \\
                u_{k-T+L-1}^{} & u_{k-T+L}^{}   & \cdots & u_{k-2}^{}   & u_{k-1}^{} &  \color{matlab_green}H_{21,k}^{}\\
                               &                &        &              &            &
                \CodeAfter
                \tikz{
                    \draw[matlab_green] (4-|6) -- (5-|6) -- (5-|2) -- (4-|2) -- (4-|6);
                    \draw[matlab_red] (1-|6) -- (4-|6) -- (4-|2) -- (1-|2) -- (1-|6);
                }
            \end{bNiceMatrix} \\
            \begin{bNiceMatrix}[first-col,last-col=6]
                \centering
                                              & u_{k-T}^{}     & u_{k-T+1}^{} & \cdots & u_{k-L-1}^{} & u_{k-L}^{}   &                                 \\
                                              & u_{k-T+1}^{}   & u_{k-T+2}^{} & \cdots & u_{k-L}^{}   & u_{k-L+1}^{} &                                 \\
                \color{matlab_red}H_{11,k}^{} & \vdots         & \vdots       &        & \vdots       & \vdots       & \color{matlab_green}H_{12,k}^{} \\
                                              & u_{k-T+L-1}^{} & u_{k-T+L}^{} & \cdots & u_{k-2}^{}   & u_{k-1}^{}   &
                \CodeAfter
                \tikz{
                    \draw[matlab_red] (2-|5) -- (5-|5) -- (5-|1) -- (2-|1) -- (2-|5);
                    \draw[matlab_green] (2-|6) -- (5-|6) -- (5-|5) -- (2-|5) -- (2-|6);
                }
            \end{bNiceMatrix}
        \end{gather*}
        \addtocounter{figure}{-1}
        \caption{$\mathcal{H}_{\scriptscriptstyle L}^{} ({u}_{\scriptstyle [k-T,k-1]}^{})$ and its correspondence with the columns (upper) and with the rows (bottom) of  $\mathcal{H}_{\scriptscriptstyle L}^{} ({u}_{\scriptstyle [k-T+1,k]}^{})$}
        \label{fig:comparison_hankel_matrices}
    \end{figure}

    \vspace{0.5\baselineskip}
    \noindent \textbf{1.}  Assuming that $u_{\scriptstyle [k-T,k-1]}$ is persistently exciting of order $L$, $\mathcal{H}_{\scriptscriptstyle L}^{} ({u}_{\scriptstyle [k-T,k-1]}^{})$ has linearly independent rows. Since the submatrix $\begin{bmatrix} H_{11,k} & H_{12,k} \end{bmatrix}  \in \mathbb{R}^{m(L-1)\times (T-L+1)}$ is also part of $\mathcal{H}_{\scriptscriptstyle L}^{} ({u}_{\scriptstyle [k-T,k-1]}^{})$, these rows are linearly independent and 
    \begin{equation*}
        \rank \begin{bmatrix} H_{11,k} & H_{12,k} \end{bmatrix} = mL-m,
    \end{equation*} 
    establishing the first point of Lemma \ref{lemma:Hankel_matrix}. Furthermore, this also establishes the lower bound 
    \begin{equation}
        \label{eq:hankel_lower_bound_1}
        \begin{aligned}
             \rank \mathcal{H}_{\scriptscriptstyle L}^{} ({u}_{\scriptstyle [k-T+1,k]}^{})  &\geq \rank \begin{bmatrix} H_{11,k} & H_{12,k} \end{bmatrix} \\
             &= mL-m.
        \end{aligned}
    \end{equation}
    
    \noindent \textbf{2.} Comparing $\mathcal{H}_{\scriptscriptstyle L}^{} ({u}_{\scriptstyle [k-T,k-1]}^{})$ to $\begin{bmatrix} H_{11,k}^{\top} & H_{21,k}^{\top} \end{bmatrix}^\top$ in Fig. \ref{fig:comparison_hankel_matrices}, the Hankel matrix has an additional column \[\begin{bmatrix} u_{k-T}^{\top} & \cdots & u_{k-T+L-1}^{\top} \end{bmatrix}^\top \in \mathbb{R}^{mL}.\] Thus, the matrix $\begin{bmatrix} H_{11,k}^{\top} & H_{21,k}^{\top} \end{bmatrix}^\top$ has, at most, one less linearly independent row than $\mathcal{H}_{\scriptscriptstyle L}^{} ({u}_{\scriptstyle [k-T,k-1]}^{})$ and therefore
    \begin{equation}
        \rank \begin{bmatrix}
            {H}_{\scriptstyle 11, k}^{^\top} & {H}_{\scriptstyle 21, k}^{^\top}
        \end{bmatrix} \geq mL-1,
    \end{equation}
    which corresponds to the lower bound on the second item.

    Similar to \eqref{eq:hankel_lower_bound_1}, the inequality above also establishes a lower bound on the rank of $\mathcal{H}_{\scriptscriptstyle L}^{} ({u}_{\scriptstyle [k-T+1,k]}^{})$ as 
    \begin{equation}
        \label{eq:hankel_lower_bound_2}
        \begin{aligned}
            \rank \mathcal{H}_{\scriptscriptstyle L}^{} ({u}_{\scriptstyle [k-T+1,k]}^{})  &\geq \rank \begin{bmatrix}
                {H}_{\scriptstyle 11, k}^{^\top} & {H}_{\scriptstyle 21, k}^{^\top}
            \end{bmatrix} \\
            &\geq mL-1.
        \end{aligned}
    \end{equation}
    This bound, however, is more strict than \eqref{eq:hankel_lower_bound_1} and corresponds to the lower bound of the third item of Lemma \ref{lemma:Hankel_matrix}.
    
    The upper bound 
    \begin{equation*}
        \rank \begin{bmatrix}
            {H}_{\scriptstyle 11, k}^{^\top} & {H}_{\scriptstyle 21, k}^{^\top}
        \end{bmatrix} \leq mL
    \end{equation*}
    corresponds to the scenario where the additional column in $\mathcal{H}_{\scriptscriptstyle L}^{} ({u}_{\scriptstyle [k-T,k-1]}^{})$ is linearly dependent on $\begin{bmatrix} H_{11,k}^{\top} & H_{21,k}^{\top} \end{bmatrix}^\top$, and therefore 
    $\rank \begin{bmatrix}
        {H}_{\scriptstyle 11, k}^{^\top} & {H}_{\scriptstyle 21, k}^{^\top}
    \end{bmatrix} = mL$.

    \vspace{0.5\baselineskip}
    \noindent \textbf{3.} The upper bound 
    \begin{equation*}
        \rank \mathcal{H}_{\scriptscriptstyle L}^{} ({u}_{\scriptstyle [k-T+1,k]}^{}) \leq mL
    \end{equation*}
    is trivial and represents the full row rank condition.

\subsection{Proof of Lemma \ref{lemma:dimensions_of_U}}
    To prove this Lemma, the Fundamental Theorem of Linear Maps must first be established.
    \begin{theorem}[Fundamental Theorem of Linear Maps \cite{axler_linear_2024}]
        \label{theorem: Fundamental Theorem of Linear Maps}
        Suppose $V$ and $W$ are vector spaces, $V$ has finite dimensions and $M \colon V \rightarrow W$ is a linear map from $V$ to $W$. Then the range of $M$ is finite-dimensional and
        \begin{equation*}
            \dim V = \dim \nullspace M + \dim \range M.
        \end{equation*}
    \end{theorem}

    The solution $z_k^*$ to \eqref{eq:loss_PE_condition_1} corresponds to the null space of 
    \begin{equation*}
        M = 
        \begin{bmatrix}
            {H}_{\scriptstyle 11, k}^{^\top} & {H}_{\scriptstyle 21, k}^{^\top}
        \end{bmatrix},
    \end{equation*}
    and its dimension can be determined using Theorem \ref{theorem: Fundamental Theorem of Linear Maps}. From Lemma \ref{lemma:Hankel_matrix}, $M$ has $mL$ columns and $mL$ or $mL-1$ independent vectors and therefore 
    \begin{equation*}
        \rank M = mL \text{ or } mL-1. 
    \end{equation*}
    These two scenarios are considered below.

    \vspace{0.5\baselineskip}
    \noindent \textbf{1.} \underline{\smash{$\rank M = mL$.}} If $M$ has $mL$ independent columns, then $\dim \range M = mL$ and Theorem \ref{theorem: Fundamental Theorem of Linear Maps} yields
    \begin{align*}
        \dim \nullspace M &= \dim V - \dim \range M \\
         &=  \dim z - \dim \range M \\ 
        &= mL - mL = 0,
    \end{align*}
    and the only solution to \eqref{eq:loss_PE_condition_1} is $z_k^* \equiv 0$.

    \vspace{0.5\baselineskip}
    \noindent \textbf{2.}  \underline{\smash{$\rank M = mL-1$.}}  If $M$ has $mL-1$ independent columns, then $\dim \range M = mL-1$ and Theorem \ref{theorem: Fundamental Theorem of Linear Maps} yields
    \begin{align*}
        \dim \nullspace M &= \dim V - \dim \range M \\
        &= mL - mL +1 = 1,
    \end{align*}
    and therefore the solution to \eqref{eq:loss_PE_condition_1} has the form
    \begin{equation}
        z_{\scriptstyle k}^{*} = \gamma z_{P,k}^{*} = , \quad z_{P,k}^{*} \neq 0,
    \end{equation}
    where $\gamma \in \mathbb{R}$ and $z_{P,k}^{*}$ is one particular solution to \eqref{eq:loss_PE_condition_1}. 

\subsection{Proof of Theorem \ref{theorem:loss_of_excitation_set}}
    To prove this theorem, the results of Lemma \ref{lemma:dimensions_of_U} are applied to \eqref{eq:loss_PE_condition_2}.

    \vspace{0.5\baselineskip}
    \noindent \textbf{1.}  For $\rank \begin{bmatrix} {H}_{\scriptstyle 11, k}^{^\top} & {H}_{\scriptstyle 21, k}^{^\top} \end{bmatrix} = mL$, Lemma \ref{lemma:dimensions_of_U} establishes that the solution to \eqref{eq:loss_PE_condition_1} is $z_k^* \equiv 0$. Replacing $z = z_k^*$ into
    \begin{equation*}
        \begin{bmatrix}
            {H}_{\scriptstyle 12, k}^{\top} & u_k^{\top}
        \end{bmatrix}^\top
        z = 0,
    \end{equation*}
   the equality is verified $\forall u_k^{} \in \mathbb{R}^m$. This means that any choice of input would guarantee excitation and therefore $\mathbb{U}_{k}^{\boldsymbol{-}} = \varnothing$. This result is expected, since the rows of $\mathcal{H}_{\scriptscriptstyle L}^{} ({u}_{\scriptstyle [k-T+1,k]}^{})$ are already independent regardless of $u_k^{}$.

    \vspace{0.5\baselineskip}
    \noindent \textbf{2.}  For $\rank \begin{bmatrix} {H}_{\scriptstyle 11, k}^{^\top} & {H}_{\scriptstyle 21, k}^{^\top} \end{bmatrix} = mL-1$, the solution to \eqref{eq:loss_PE_condition_1} is 
    \begin{equation*}
        z_{\scriptstyle k}^{*} = \gamma z_{P,k}^{*}, \quad z_{P,k}^{*} \neq 0, \quad \gamma \in \mathbb{R}.
    \end{equation*}
    Defining the partition
    \begin{equation*}
            z_{P,k}^{*} = \begin{bNiceMatrix}
                b_{\scriptstyle k}^{} \\ a_{\scriptstyle k}^{}
            \end{bNiceMatrix}, \quad a_{k}^{} \in \mathbb{R}^{m}, \quad b_{\scriptstyle k}^{} \in \mathbb{R}^{mL-m},
        \end{equation*}
    and replacing $z = z_k^*$ into \eqref{eq:loss_PE_condition_2}, the input $u^{\boldsymbol{-}}$ that causes loss of excitation can be determined as 
    \begin{equation}
        a_{k}^\top u^{\boldsymbol{-}} + b_{k}^\top H_{12, k} = 0,
    \end{equation}
    which corresponds to the equation of a hyperplane. $\mathbb{U}_{k}^{\boldsymbol{-}}$ is then given by
    \begin{equation}
        \mathbb{U}_k^{\boldsymbol{-}} \coloneq \{ u_{\scriptstyle k}^{} \in \mathbb{R}^m \mid a_{\scriptstyle k}^\top u_{\scriptstyle k}^{} +b_{k}^\top H_{12, k} = 0\}.
    \end{equation}
    
    Theorem \ref{theorem: Fundamental Theorem of Linear Maps} can be used to determine the dimension of $\mathbb{U}_k^{\boldsymbol{-}}$. Given that $T = a_{k}^\top$ has $m$ columns and has one row, Theorem \ref{theorem: Fundamental Theorem of Linear Maps} yields 
    \begin{align*}
        \dim \nullspace  T &= \dim V - \dim \range T \\
        \dim \nullspace a_{k}^\top  &=  \dim u_{\scriptstyle k}^{} - \dim \range  a_{k}^\top \\ 
        &= m - 1 ,
    \end{align*}
    and therefore the dimension of $\mathbb{U}_k^{\boldsymbol{-}}$ is $m-1$ and $\mathbb{U}_k^{\boldsymbol{-}} \in \mathbb{R}^{m-1}.$

\addtolength{\textheight}{-12cm}   

\end{document}